\documentclass[english,tightenlines,a4paper,twocolumn]{revtex4}

\usepackage[pdftex,dvips,final]{graphicx}

\begin{document}

\title{CdSe quantum dots in ZnSe nanowires as\\efficient source for single photons up to 220~K}

\author{Thomas Aichele}
\email[Corresponding author. E-mail: ]{thomas.aichele@physik.hu-berlin.de} \altaffiliation{present
address: Institut f{\"u}r Physik, Humboldt Universit{\"a}t zu Berlin, Hausvogteiplatz 5--7, 10117
Berlin, Germany}
\author{Adrien Tribu}
\author{Gregory~Sallen}
\author{Juanita Bocquel}
\author{Edith Bellet-Amalric}
\author{Catherine Bougerol}
\author{Jean-Philippe Poizat}
\author{Kuntheak Kheng}
\author{R\'egis Andr\'e}
\author{Serge Tatarenko}
\address{Nanophysics and Semiconductors Group, CEA/CNRS/University J. Fourier, 17 rue des Martyrs, 38000 Grenoble,
France}

\date{\today}

\begin{abstract}
ZnSe nanowire heterostructures were grown by molecular beam epitaxy in the vapour-liquid-solid
growth mode assisted by gold catalysts. Size, shape and crystal structure are found to strongly
depend on the growth conditions. Both, zinc-blende and wurtzite crystal structures are observed
using transmission electron microscopy. At low growth temperature, cone-shaped nano-needles are
formed. For higher growth temperature, the nanowires are uniform and have a high aspect ratio with
sizes of 1--2~$\mu$m in length and 20--50~nm in width as observed by scanning electron microscopy.
Growing a nanowire on top of a nano-needle allows us to obtain very narrow nanorods with a diameter
less than 10 nm and a low density of stacking fault defects. These results allow us the insertion
of CdSe quantum dots in a ZnSe nanowire. An efficient photon anti-bunching was observed up to
220~K, demonstrating a high-temperature single-photon source.
\end{abstract}

\maketitle

An appealing application for semiconductor nanowires (NWs) is the inclusion of quantum dots (QDs)
into the NW. Due to the narrow lateral size, NW QD heterostructures can be directly grown on
defined positions and without the necessity of self-assembly. This is especially important for for
II-VI materials, where self-assembled QD formation occurs only within narrow windows of growth
conditions \cite{RobinAPL}. Recently, II-VI compound semiconductor NWs have been synthesized by
Au-catalysed metal-organic chemical vapour deposition (MOCVD) and molecular-beam epitaxy (MBE)
methods \cite{NW-MOCVD,NW-MBE,Colli}. QD devices can be utilized as emitters for the effective and
controlled generation of single-photon states. Single-photon emission from a GaAsP/GaP NW QD was
reported before at cryogenic temperature in ref. \cite{Zwiller_NW_AB}. High-temperature experiments
from individual Stransky-Krastanov (SK) grown QDs were reported from CdSe/ZnSe QDs \cite{Sebald}
and from GaN/AlN QDs \cite{Kako}. Both experiments showed photon anti-bunching up to 200~K with
normalized dip values of 81\% and 53\%, respectively. Other systems have demonstrated room
temperature single photon emission: nanocrystals \cite{MichlerNC} have the drawback that they
suffer from blinking effect \cite{BrokmannBlinking}; colour centres in diamond
\cite{Grangier,Kurtsiefer} have shown very reliable operation but with a very broad spectrum.
Anyways, neither nanocrystals nor colour centres in diamond offer the possibility of electrical
excitation, which is a very realistic and very promising perspective for semiconducting nanowires
\cite{MinotNWLED}.

In this paper, we report MBE growth of ZnSe NWs on Si and GaAs substrates and insertion of CdSe QDs
in these NWs. The growth process is based on the Au-catalyzed VLS method. The morphology of the NWs
depending on the growth parameters was examined by scanning electron microscopy (SEM) and
high-resolution transmission electron microscopy (HRTEM). We found two different growth regimes,
resulting in either narrow and uniform NWs, or cone-shape nano-needles. When combining these growth
regimes, NWs with a very low density of defects can be achieved. As an important application of
these NW QD structures, we present the generation of single photons from a single QD with a deep
anti-bunching of the photon correlation even up to high a temperature of 220~K.

\begin{figure}%
\includegraphics*[width=\linewidth]{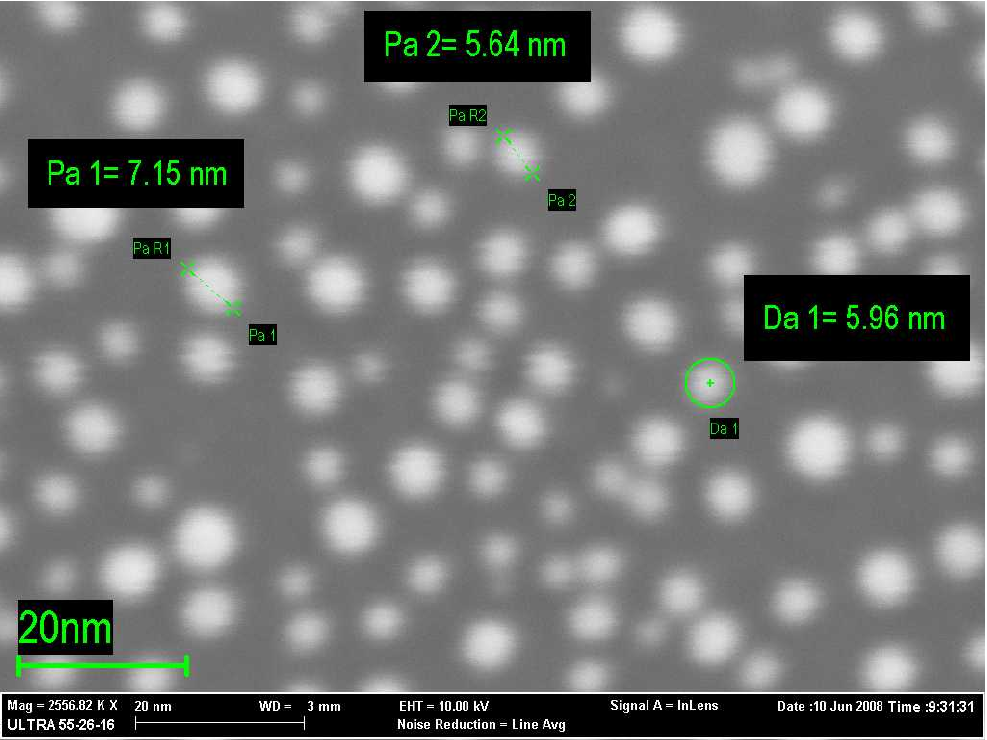}
\caption{SEM image of a 0.5 nm thick gold layer dewetted on a Si(100) substrate. The numbers
indicate diameters of a few selected gold particles.}
\end{figure}

Prior to the growth, the substrates, (100)-, (111)-GaAs and (100)-Si, were annealed in an
ultra-high vacuum (UHV) chamber to 580$^\circ$ C in order to degas the surfaces.
Additionally, the GaAs surface deoxidizes at this temperature, while the Si substrate
remains oxidized. A thin Au layer (0.2--1 nm thick) was then deposited on the substrate
at room temperature by e-beam metal deposition. The samples were then introduced in a
II-VI MBE chamber. The transfer between the MBE and metal deposition chamber were
performed under UHV. In order to generate Au nanoparticles, the gold film was dewetted by
annealing the substrate to 450$^\circ$ C for 10 min. At this temperature, the gold forms
nanoparticles, as observed by SEM and AFM (Figure 1(a)). Next, growth of ZnSe wires at
different sample temperatures was performed. All samples were grown by solid-source MBE.
The growth parameters like growth temperature, Zn:Se flux ratio or gold thickness were
varied independently. The growth rate is in the 0.5 nm/s range.

\begin{figure}%
\includegraphics*[width=\linewidth]{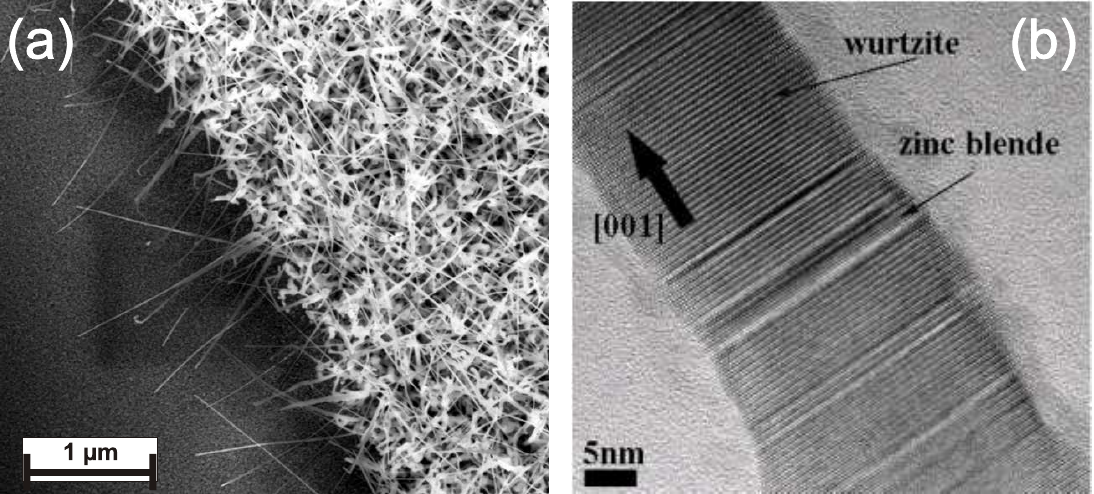}
\caption{ZnSe NW grown at 350$^\circ$ C in excess of Se. (a) SEM image of the as-grown sample; (b)
TEM image of a single NW.}
\end{figure}

First we studied the effect of the sample temperature on the NW structure, while the beam
pressure where kept constant (Zn (Se) flux: 2.5 (7.5) $\times10^{-7}$ torr). When growing
at a substrate temperature within 350-450$^\circ$ C, a dense carpet of uniform NWs covers
the substrate (fig. 2(a). They have a diameter of about 20--50 nm and a length of 1--2
$\mu$m. The structure is predominantly a wurtzite structure with [0001]-axis as growth
direction. As seen in the TEM image in fig 2(b), stacking faults due to zinc blende
insertions are repeatedly observed along the NW. Additionally to the NWs, the substrate
is covered with highly irregular nano-structures. The formation of stacking faults and
irregular nano-structures are explained by non-ideal growth conditions at the initial
stages of the growth process. Possible reasons are the presence of non-uniform gold
agglomerations instead of small gold beads and the insertion of impurities during the
gold deposition process. The presence of both wurtzite and Zn-blende shows, that at the
utilized growth conditions both phases are allowed.

\begin{figure}%
\includegraphics*[width=\linewidth]{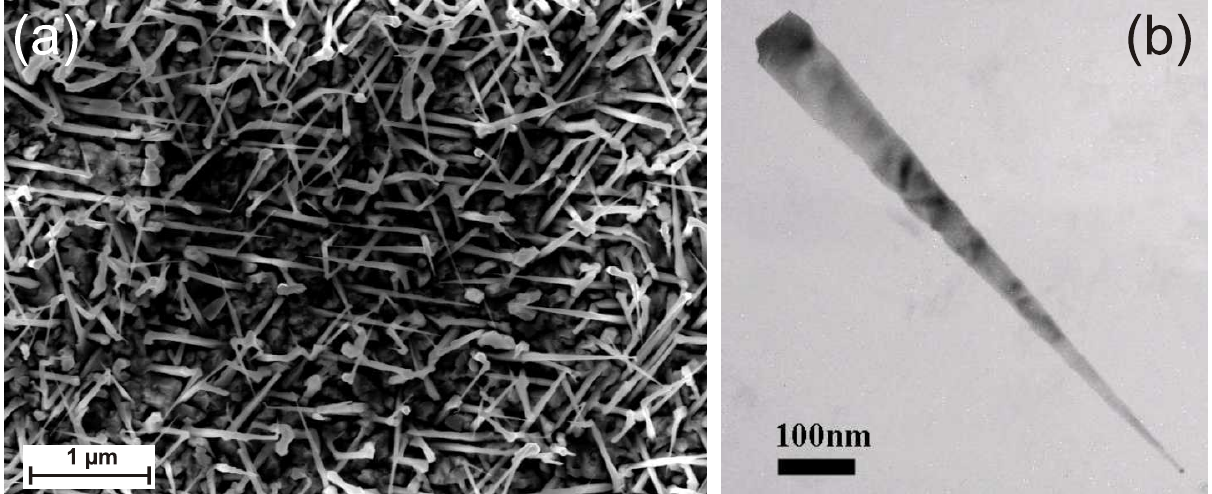}
\caption{ZnSe nano-needle grown at 300$^\circ$ C in excess of Se. (a) SEM image of the as-grown
sample; (b) TEM image of a single NW.}
\end{figure}

When growing at low temperature (300$^\circ$ C), HRTEM and SEM images reveal the formation of
nano-needles with a wide base (80 nm diameter) and a sharp tip (5--10 nm diameter) covered by a
gold particle as seen in fig. 3. Similar results are obtained at a growth temperature of
450$^\circ$ C but with inverted Zn:Se flux ratio. The structure is mainly a wurtzite type with
[11-20]-axis as growth direction. Towards the base, the structures are again repeatedly intersected
by zinc blende domains, while the tip has a pure wurtzite structure. The formation of nano-needles
instead of NWs is well accounted for by the slower adatom mobility expected at low temperature,
which promotes nucleation on the sidewalls before reaching the gold catalyst at the nano-needle
tip. A similar idea was proposed in ref. \cite{Colli} for the formation of nanosaws. In contrast to
the NWs, the defect planes are here disoriented with respect to the nano-needle axis. It seems that
this disorientation hinders the propagation of defects in the growth direction, especially for
lower diameters. Defects zones are rapidly blocked on the side walls, providing a higher structural
quality towards the nano-needle tip. Interestingly the nanowire structure and shape do depend only
little on the underlying substrate.

\begin{figure}%
\includegraphics*[width=\linewidth]{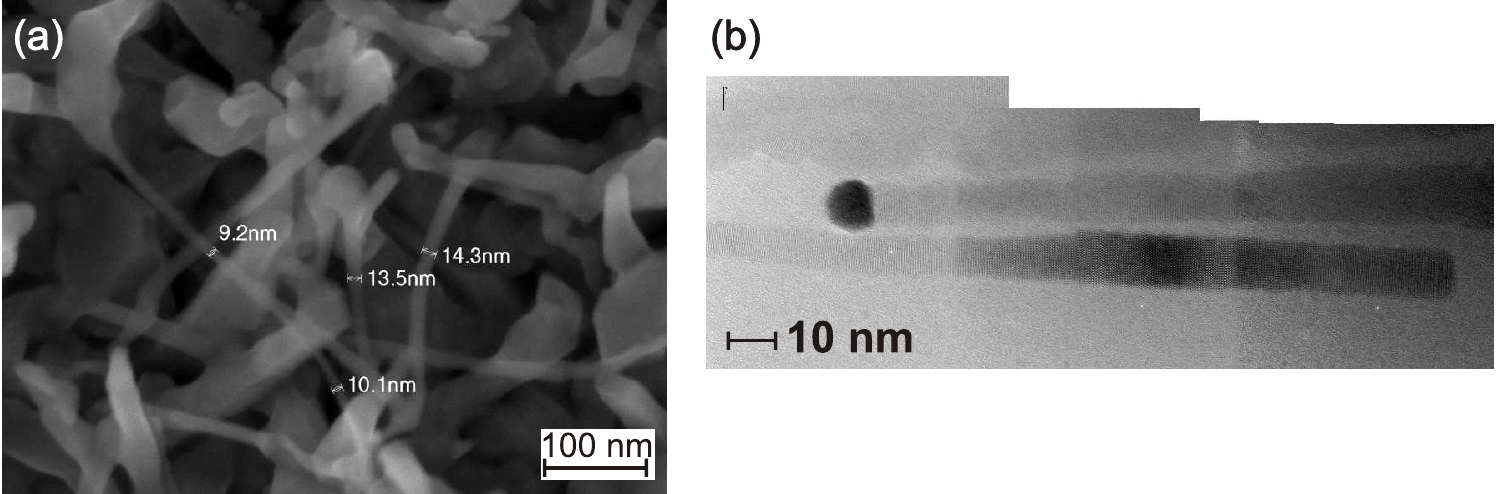}
\caption{Two-step growth of a ZnSe NW on a nano-needle. (a) SEM image of the as-grown sample; (b)
collage of TEM images of two close-by NWs.}
\end{figure}

The observation of a decreasing defect density from the base towards the top in the nano-needles
motivated us to modify the growth recipe in the following way: In the first part, the sample is
grown with excess of Zn for 30~min, leading to the formation of nano-needles. Next, the Zn- and
Se-flux was inverted and NWs were grown for another 30 min on top of the nano-needles. Thus, the
growth at the side-walls was aborted and re-growth started on defect-free and strain-relaxed
nano-needle tips, where the high structural quality of the crystal lattice can be preserved along
the narrow NW that is now formed in this second growth step. Fig. 4 shows results obtained from
this sample. The structures have a broad base that tapers after a few ten nanometers to thin NWs
with thickness of 10--15~nm. When studying single NWs by HRTEM, we see that the stacking faults
indeed strongly reduce towards the thin part of the NW.

\begin{figure}%
\includegraphics*[width=\linewidth]{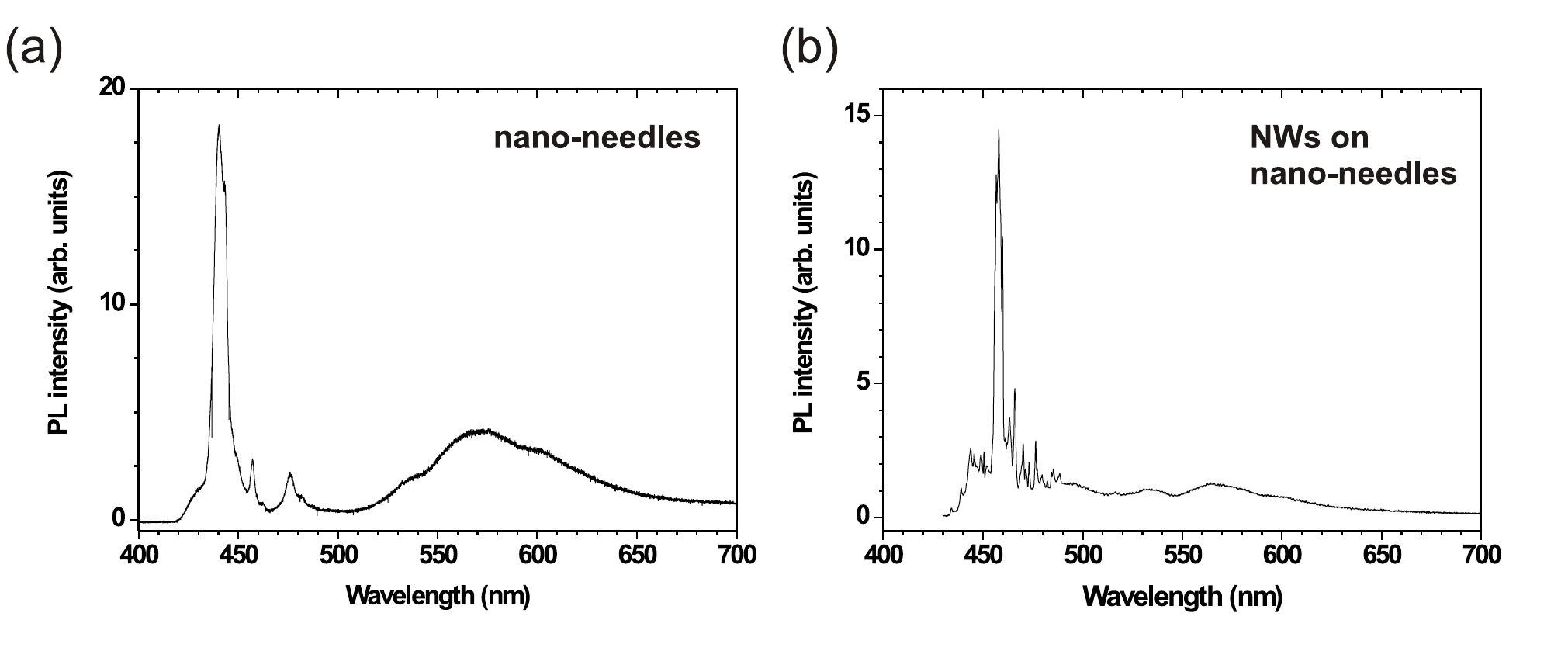}
\caption{Ensemble spectrum from (a) the nano-needle sample in fig. 3, and (b) the combined
NW/nano-needle sample in fig. 4.}
\end{figure}

The suppression of such stacking faults becomes important, when observing the spectral
properties of a single NW QD. The spectra were taken at a sample temperature of 5~K. The
samples were excited by a cw laser at 405~nm via a microscope objective. Fig. 5(a) shows
the spectra from an ensemble of nano-needles. The peak at 440~nm belongs to ZnSe bandedge
emission. This is mostly due to bulk ZnSe that is grown in parallel to the nano-needles
at the sample floor. Single nano-needles and NWs (not shown here) only show a very weak
contribution of the ZnSe bandedge. More strikingly, we observed a broad emission within
500--600 nm. This can be attributed to excitons localized at the defect zones in the
nano-needle \cite{Philipose2}. In contrast, for the combined NW/nano-needle structure,
fig. 5(b), this emission is mostly suppressed leaving behind a only a small bunch of
spectral lines between 450--500~nm. These are presumably from the broader base of these
structures. When isolating single structures (see below), it will most likely break at
the narrow NW part, which contains much less defects, so that we find single NWs, in
which this broad emission is completely suppressed.

With these conditions, it finally becomes possible to grow and study single NW samples
with an inserted CdSe QD. In order to prepare QDs, a small region of CdSe has been
inserted in this high quality part of the ZnSe NW. This is done by interrupting the ZnSe
growth after 30 min and changing to CdSe for 30~s. Next, the ZnSe growth is continued for
another 15~min. The diameter size is of the order of the bulk exciton Bohr diameter for
CdSe (11 nm). This means that the carriers in the CdSe QD are in the strong confinement
regime. For the study of single NWs, the sample is put in a methanol ultra-sonic bath for
30 s, in which some NW broke off the substrate into the solution. This process allows
also to detach mainly the high quality part of the NW from the nano-needles where many
stacking faults are present. Droplets of this solution were next placed on a new
substrate, leaving behind a low density of individual NWs. Metal markers have been made
on the substrate using optical lithography technique in order to locate the NWs.

\begin{figure}%
\includegraphics*[width=\linewidth]{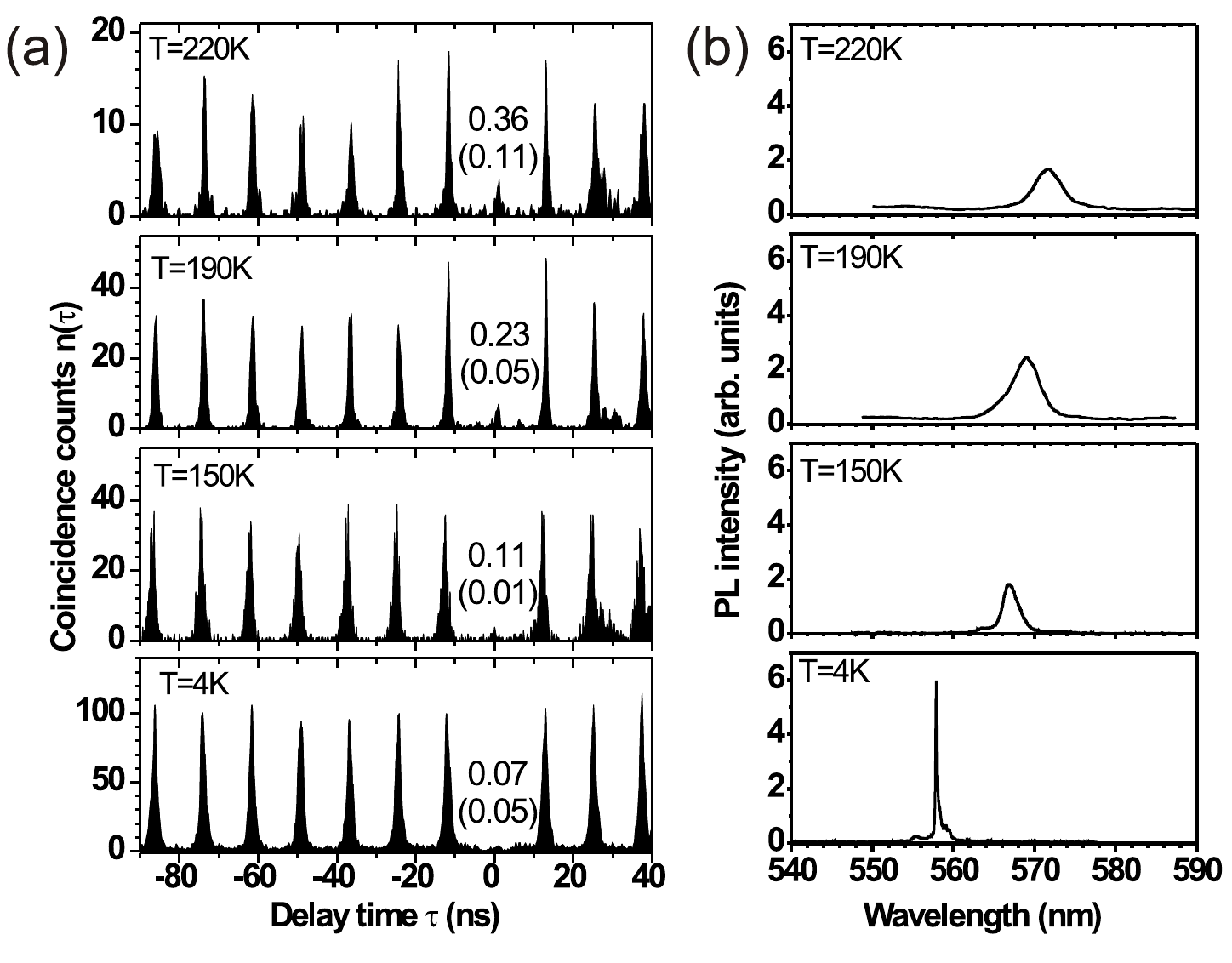}
\caption{(a) Second-order correlation taken under pulsed excitation at temperatures 4 and 220~K,
respectively. The numbers in the graphs are measured values of $g^{(2)}(0)$ (i.e. the area under
the peak at $\tau=0$ relative to the peaks at $|\tau|>0$). The numbers in parentheses are the
reconstructed values of the {\em pure} spectral lines taking into account a spectral background.
(b) Spectra from the same NW QD taken in the same experimental run as in (a).}
\end{figure}

Fig. 6(b) shows the spectra from a NW QD at different temperatures. The spectral line can be
attributed to a trion transition. With higher excitation intensity, also an exciton and biexciton
transition become visible. Above 150~K, the spectral lines significantly broaden. We found that the
light emission is highly polarized with a contrast of 80--90\% (while exciting the NW with
circularly polarized laser light). This striking polarization anisotropy of absorption can be
explained by the dielectric contrast between the NW material and the surrounding environment, and
the orientation of the dipole within the QD \cite{LieberPolNW,NiquetPolNW}.

We have carried out photon correlation measurements using a Hanbury Brown and Twiss
setup. The sample (few NWs on a silicon substrate with micro-markers) was mounted in a
He-flow cryostat. The QD's were optically excited with a frequency doubled Ti:sapphire
laser emitting 200-fs pulses with a repetition rate of 80 MHz. The laser was focused on
the sample by using a microscope objective and the luminescence was collected with the
same objective. The QD emission passed through a spectrometer and is then sent to a
Hanbury Brown-Twiss correlator based on two silicon avalanche photodiodes (APDs) and a
coincidence counter.

The graphs in fig. 6(a) are the raw histograms of measured coincidences without any
correction for background count events. The area under each peak at time $\tau=0$, was
normalized with respect to the average area under the peaks at $|\tau|>0$. Each peak area
was calculated by integrating the coincidences within 12-ns windows (repetition time of
the excitation laser). The correlation functions were taken at different temperatures
between 4~K and 220~K. At 4~K, the peak at $\tau=0$ is suppressed to a normalized value
of 7\%, showing the high quality of the single-photon generation. With increasing
temperature, this value only slightly increases to finally reach 36\% at 220~K. This
value is far below 50\%, the emitted light field is thus clearly distinguished from
states with 2 or more photons. Thus, even without correcting for background events, these
emitters can be directly used as a high-quality single-photon device even when operating
at high temperature, with a strongly suppressed probability for two-photon events. In the
correlation measurement obtained at 220~K, a low resolution grating was used in the
spectrometer so that all the photons coming from the broader line could be counted. This
broader spectral window of integration leads to larger background, which is the main
origin for the rise of the $\tau=0$-peak above 150~K. In contrast to SK QDs which often
grow with a high density on the substrate, the density of NWs in the microscope focus was
much smaller and can be even reduced to only one within the microscope focus, which
avoids contributions from neighbouring emitters that spectrally overlap with the
transition under observation.

In summary, we have performed optical studies on single CdSe QDs in ZnSe NWs. The NWs
were developed by MBE in a two-step growth recipe, where narrow, mostly defect-free NWs
are grown on top of broader, cone-shaped NWs. The single-NW PL features narrow, isolated
spectral lines. When filtering individual transitions, non-classical single-photon
statistics were retrieved, indicated by strong anti-bunching, where the {\em raw}
correlation function was reduced down to a normalized value of 7\%. This behaviour
remains even up to a temperature of 220~K, where this correlation peak is only slightly
increased to 36\%. For non-blinking QDs, this is the highest reported temperature for
single-photon emission and for an anti-bunching-dip below 50\%. At this temperature,
Peltier cooling becomes an alternative to liquid helium or nitrogen cooling. Together
with the possibility of integrating NWs into electro-optical circuits \cite{MinotNWLED},
these emitters become an interesting candidate for developing compact, stable and
cost-efficient quantum devices operating near room temperature.

\section*{Acknowledgements}

T.A. acknowledges support by Deutscher Akademischer Austauschdienst (DAAD). Part of this work was
supported by European project QAP (Contract No. 15848).

\end{document}